\documentclass{aa}
\usepackage{txfonts}
\usepackage{graphicx}
\usepackage{natbib} 
\bibpunct{(}{)}{;}{a}{}{,} 

\begin{document}
\newcommand{\Planck}{{\sc Planck}}
\newcommand{\fk}{f_{\rm k}}
\newcommand{\deltaTWN}{\delta T_{\rm sky(ref)}^{\rm WN}}
\newcommand{\deltaTWNi}{\delta T_{{\rm sky(ref)},i}^{\rm WN}}
\newcommand{\sigmaWN}{\sigma_{\rm sky(ref)}^{\rm WN}}
\newcommand{\sigmaskyWN}{\sigma_{\rm sky}^{\rm WN}}
\newcommand{\sigmarefWN}{\sigma_{\rm ref}^{\rm WN}}
\newcommand{\sigmaskyN}{\sigma_{\rm sky}^{N}}
\newcommand{\sigmaskyrefN}{\sigma_{\rm sky(ref)}^{N}}
\newcommand{\sigmarefN}{\sigma_{\rm ref}^{N}}
\newcommand{\Vskyi}{V_{{\rm sky,}i}}
\newcommand{\Vskyrefi}{V_{{\rm sky(ref),}i}}
\newcommand{\Vskyref}{V_{\rm sky(ref)}}
\newcommand{\Vrefi}{V_{{\rm ref,}i}}
\newcommand{\Tsky}{T_{\rm sky}}
\newcommand{\Vsky}{V_{\rm sky}}
\newcommand{\Tni}{T_{{\rm n,}i}}
\newcommand{\Tref}{T_{\rm ref}}
\newcommand{\Vref}{V_{\rm ref}}
\newcommand{\Ttsky}{\tilde T_{\rm sky}}
\newcommand{\Ttref}{\tilde T_{\rm ref}}
\newcommand{\Ttskyref}{\tilde T_{\rm sky(ref)}}
\newcommand{\Tn}{T_{\rm n}}
\newcommand{\Tphys}{T_{\rm phys}}
\newcommand{\Tnfe}{T_{\rm n_{\rm FE}}}
\newcommand{\Tnbe}{T_{\rm n_{\rm BE}}}
\newcommand{\Ns}{N_{\rm s}}
\newcommand{\Tsys}{T_{\rm sys}}
\newcommand{\Lsky}{L_{\rm sky}}
\newcommand{\Lref}{L_{\rm ref}}
\newcommand{\Gfe}{G_{\rm FE}}
\newcommand{\Gbe}{G_{\rm BE}}

\title{Offset balancing in pseudo-correlation radiometers for CMB measurements}

\author{
    Aniello Mennella \inst{1}\and
    Marco Bersanelli \inst{2}\and
    Michael Seiffert \inst{3}\and
    Danielle Kettle  \inst{4}\and
    Neil Roddis      \inst{4}\and
    Althea Wilkinson \inst{4}\and
    Peter Meinhold   \inst{5}
} \offprints{A. Mennella}

\institute{
    IASF-CNR, Sezione di Milano, Via Bassini 15, 20133 Milan,
    Italy\and
    Universit\`a degli Studi di Milano, Via Celoria 16, 20133 Milan,
    Italy\and
    Jet Propulsion Laboratory, California Institute of Technology,
    Pasadena, CA 91109, USA\and
    Jodrell Bank Observatory, Jodrell Bank, Macclesfield, Cheshire,
    SK11 9DL, UK\and
    University of California at Santa Barbara, Physics Department, Santa Barbara, CA 93106,
    USA
}

\date{Received 26 March 2003 / Accepted 8 July 2003}

\abstract{
        Radiometeric CMB measurements need to be highly stable and
        this stability is best obtained with differential receivers.
        The residual 1/$f$ noise in the differential output
        is strongly dependent on the radiometer
        input offset which can be cancelled using various balancing strategies.
        In this paper we
        discuss a software method implemented in the
        \Planck-LFI pseudo-correlation receivers which uses a tunable
        {\em gain modulation factor}, $r$, in the sky-load difference.
        Numerical simulations and experimental data
        show how proper tuning of the parameter $r$ ensures a very stable differential
        output with knee frequencies of the order of few mHz. Various approaches
        to calculate
        $r$ using the radiometer total power data are discussed with some
        examples relevant to \Planck-LFI.
        Although the paper focuses on pseudo-correlation receivers and the
        examples are relative to \Planck-LFI, the proposed method 
        and its analysis is general and can be applied to a large class of 
        differential radiometric receivers.
}

\authorrunning{A. Mennella et al.}

\maketitle

\section{Introduction}
\label{sec:introduction}
        The dramatic progress achieved in the past decade in Cosmic Microwave Background
        (CMB) observations, particularly in anisotropy experiments,
        is strongly correlated with the remarkable improvements obtained in
        microwave and sub-millimetre detector technology, as well as in cryogenic
        technology \citep[for a recent review see, e.g.,][]{Bersanelli:2002}.
        Single-detector sensitivities of 0.1 - 0.4 mK Hz$^{-1/2}$ have been demonstrated
        for cryogenic operation of Indium Phosphide HEMT (High Electron Mobility
        Transistors)
        amplifiers (typically cooled at $\sim 20$ K) in the range 25-100 GHz.
        At higher frequencies, sensitivities at a level ${\rm NEP} \sim 1 \times
        10^{-17}$~W$\times$Hz$^{-1/2}$ have been achieved by spider web bolometers
        cooled to
        $\sim 0.1$ K \citep[see, e.g.,][]{Lamarre:1997}. Moderate-size arrays of such
        detectors can today produce
        high resolution full-sky maps of the CMB with a high signal-to-noise ratio.
        These ultra-sensitive systems impose stability requirements that are
        proportionally stringent
        and call for highly optimised instrument design.
        In particular, the instrument needs to be immune at $\mu$K level from
        the effect of parasitic signals introduced by non-idealities in the system,
        which would
        propagate as systematic errors in the final CMB maps
        \citep{Mennella:2002}.

        In coherent radiometeric systems one of the major concerns is
        the intrinsic instability due to gain and noise temperature fluctuations
        of the amplifiers themselves,
        typically well represented by a $1/f$-type noise spectrum.
        Differential receivers,
        such as the Dicke-switched scheme, reduce the impact of amplifier instabilities
        in the measured signal with a fast (typically $\sim 100$ Hz) switch between
        the sky input port and a stable reference, sometimes given by another
        horn pointed at the sky. Dicke-type receivers have a long history in CMB
        observations
        and were successfully employed in the COBE-DMR instrument that first detected
        CMB anisotropies \citep{Smoot:1990,Smoot:1992}.

        In recent years, a scheme called ``pseudo-correlation'' radiometer has been
        introduced to improve over the classical Dicke scheme. As it will be
        discussed in detail, this design has a two-port front-end that
        allows a continuous comparison between the
        sky signal $\Tsky$ and a stable reference signal $\Tref$,
        improving the sensitivity by a factor $\sqrt{2}$
        over a Dicke radiometer. In addition, fast (few kHz) phase
        switching provides immunity from back-end fluctuations.
        Different versions of pseudo-correlation designs are being
        used for the second and third generation of space-based
        radiometric instruments for CMB anisotropy: NASA's Wilkinson Microwave Anisotropy Probe
        (WMAP) and the Low Frequency Instrument (LFI) on board ESA's Planck mission.

        In principle, a perfectly balanced pseudo-correlation radiometer
        is completely free from 1/$f$ effects.
        Residual sensitivity to $1/f$ noise, as well as to other systematic effects,
        is proportional to the input offset $\Delta T_{\rm off} \equiv \Tref - \Tsky$
        at the level of the first hybrid coupler in the radiometer front-end.
        In practice full balance is not achievable, so that a key instrument
        design objective is to minimise $\Delta T_{\rm off}$.   

        In the case of the WMAP instrument, the radiometers directly measure temperature
        differences between sky signals from two widely separated regions of the sky
        \citep{Jarosik:2003,Bennett:2003}.
        This is accomplished with a symmetric back-to-back double-telescope system,
        and pairs of feeds which provide the two inputs to the pseudo-correlation 
        front-end.
        In this scheme, the contribution to the offset from external signals is only of
        few mK
        (dominated by the CMB dipole of Galactic plane at low frequencies).
        The offset (of order $\Delta T_{\rm off} <$1~K) is dominated by second-order
        instrument asymmetries.

        The LFI radiometers, instead, measure differences between the sky, $\Tsky$, and
        a stable
        internal cryogenic reference load ($\Tref$) cooled at about 4~K by
        the pre-cooling stage of the High Frequency
        Instrument (HFI) bolometer array
        in the Planck focal plane. This introduces an offset
        of order $\Delta T_{\rm off} \sim 2$ to 3~K.  The effects of this offset,
        however,
        are compensated by introducing a
        ``gain modulation factor'', $r$, which balances the output in the
        on-board signal processing
        \citep{Bersanelli:1995,Seiffert:2002}.
        Experimental results from advanced LFI prototypes
        \citep{Meinhold:1998, Tuovinen:2000},
        as well as analytical calculations,
        show that great immunity from $1/f$ effects
        can be obtained when the value of $r$ is accurately selected.
        
        Two different approaches can be used to select the value of $r$:
        it can either be controlled in hardware
        by adjusting (in principle in real-time)
        a variable gain to achieve the null-output condition;
        or it can be set as a controllable software parameter,
        giving more flexibility to the system. The latter approach was adopted for the
        LFI.
        
        As we shall see in more detail, the best estimate of
        $r$ can be obtained based on the radiometer data themselves,
        collected in a raw, undifferenced form.
        In a space application this may impose
        non-trivial demands on the telemetry rate.
        If the data available to calculate $r$ are limited by telemetry constraints,
        this may limit the accuracy of the determination of $r$ and therefore
        limit the instrument stability.
        
        A detailed analytical study of the
        impact of non-idealities in the LFI radiometer, in
        particular $1/f$ noise effects, has been carried out
        in a previous work \citep{Seiffert:2002}.
        In this paper we discuss in detail the required and obtainable
        accuracy of $r$, and its effect on the measurement quality.
        Although we focus on offset balancing in pseudo-correlation radiometers
        and discuss some examples in the context of \Planck-LFI,
        the concepts and formalism discussed in our paper are general
        and applicable to any switched radiometer
        like unbalanced Dicke receivers.
        
        After a short description of the radiometer concept,
        we discuss the issue of offset balancing and
        derive the optimal $r$ for $1/f$
        noise minimisation (Sect.~\ref{sec:offset_balancing}), demonstrating its
        effectiveness by applying it
        to representative simulated data streams. In addition we
        discuss the impact of the choice of $r$ on other systematic
        effects and derive the required accuracy
        on the calculation of $r$.
        In Sect.~\ref{sec:calculation_of_r} we present and compare various approaches
        for calculating
        $r$ and show how these can be used in the context of space experiments; an example
        of the application of this offset-balancing technique to laboratory radiometer
        data is presented
        and discussed in Sect~\ref{sec:example_experimental}.
        Finally, in Sect.~\ref{sec:impact_of_systematic_effects} we discuss the impact
        of systematic effects
        of instrumental and astrophysical origin on the calculation of $r$.

\section{Pseudo-correlation differential radiometers: basic concepts}
\label{sec:pseudo_corr_basic}

        The Dicke switched radiometer represents a
        typical implementation of a differential receiver. In its simplest form
        a switch located in the front-end commutes rapidly between the sky and the
        reference
        horns so that a sequence of sky-load
        signals is detected and differenced at the output of the amplification chain;
        the radiometer
        sensitivity per unit integration time is given by:
        \begin{equation}
                \Delta T_{\rm rms} = 2\frac{T_{\rm sys}}{\sqrt{\beta}},
        \end{equation}
        where $T_{\rm sys} = T_{\rm sky} + T_{\rm noise}$ is the system temperature,
        $T_{\rm noise}$ represents the radiometer noise temperature and $\beta$ is the
        bandwidth.
        
        If the switching frequency is sufficiently high
        so that the gain can be considered nearly constant in a sky-load cycle, then
        the susceptibility to gain fluctuations is very much improved:
        \begin{equation}
                \frac{\Delta T_{\rm Gain}}{T_{\rm sys}} =
                \frac{\Delta G}{G}\frac{T_{\rm sky}-T_{\rm ref}}{T_{\rm sys}},
                \label{eq:dicke}
        \end{equation}
        where $\Delta T_{\rm Gain}$ represents the rms signal variation caused by gain
        fluctuations
        and $T_{\rm sky}$, $T_{\rm ref}$ represent
        the sky and reference load temperatures, respectively.
        
        Although this scheme is effective in reducing 1/$f$ noise in the final
        measurements, the
        presence of a lossy active component in the front-end increases the radiometer
        noise and may introduce additional 1/$f$ components. These
        limitations, however, can be overcome with a modified version of the Dicke
        scheme,
        the {\em pseudo-correlation radiometer}, that has been recently adopted for the
        WMAP and \Planck-LFI instruments.
        
\begin{figure}[here]
                \begin{center}
                        \resizebox{8.5 cm}{!}{\includegraphics{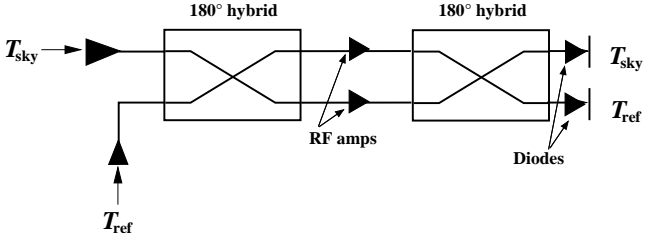}}
                \end{center}
                \caption{
                        A pseudo-correlation radiometer in its simplest form.
                        This scheme allows a differential measurement without requiring
                        an active switch before the first RF gain stage.
                }
                \label{fig:pseudo_simple}
        \end{figure}

        In Fig.~\ref{fig:pseudo_simple} we show a schematic of a pseudo-correlation
        radiometer
        in its simplest form. The sky and reference signals are summed by a front-end
        180$^\circ$
        hybrid, amplified by two parallel amplification chains and then separated
        by a second hybrid, detected and differenced. In this way the 1/$f$ noise
        from the RF amplification stages is the same in both signals, so that they can
        be eliminated at first order by differencing.
        
\begin{figure}[here]
                \begin{center}
                        \resizebox{8.5 cm}{!}{\includegraphics{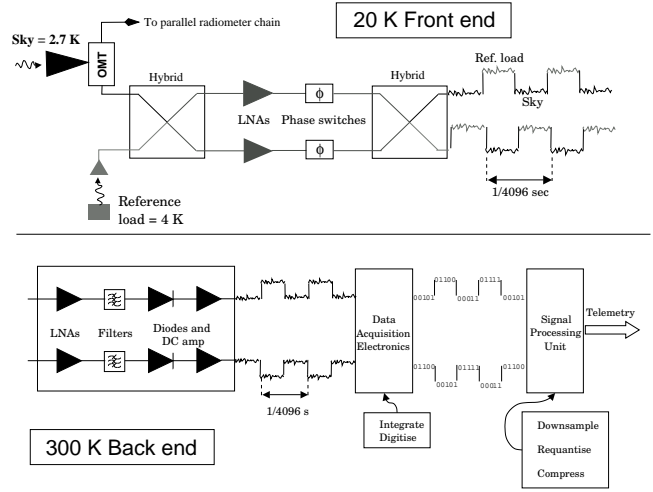}}
                \end{center}
                \caption{
                        Baseline LFI pseudo-correlation radiometer.
                        One of the two 180$^\circ$ phase switches in the front end switches rapidly the
                        phase of the propagating signal, thus producing a sequence of sky-load signals
                        at the output of the second hybrid.
                        In the warm back-end the signal are further amplified, detected and digitised
                        before telemetry.
                }
                \label{fig:LFI_scheme}
        \end{figure}

        To reduce the effect of instabilities in the back-end electronics a fast
        switching
        between the two inputs is implemented in the two radiometer legs;  as
        an example of application of the switched pseudo-correlation scheme,
        Fig.~\ref{fig:LFI_scheme} shows a schematic of the radiometers used in the {\sc
        Planck}-LFI
        instrument.
        
        In the front-end part (see top part of figure) the radiation entering the
        feed-horn is separated by an OrthoMode Transducer (OMT) into two
        perpendicular linearly polarised components that propagate
        independently through two parallel radiometers. In each
        radiometer, the sky signal and the signal from a stable reference
        load at $\sim$4~K are coupled to cryogenic low-noise HEMT
        amplifiers via a $180^\circ$
        hybrid. One of the two signals then runs through a switch that
        applies a phase shift which oscillates between 0 and $180^\circ$ at a
        frequency of 4096~Hz\footnote{
          This frequency corresponds to a full
          0-180$^\circ$ cycle, so that the rate between two subsequent switch states
          is 8192~Hz
        } (the second phase switch is present for
        symmetry on the second radiometer leg but it does not
        introduce any phase shift). The signals are then
        recombined by a second 180$^\circ$ hybrid coupler, producing a sequence of
        sky-load signals at the output alternating at twice the frequency of the phase
        switch.
        
        In the back-end of each radiometer (see bottom part of
        Fig.~\ref{fig:LFI_scheme}) the RF signals are further
        amplified, filtered by a low-pass filter and then detected. 
        After detection the sky and reference load signals are integrated and
        digitised before sending to ground. Note that the current \Planck-LFI available 
        telemetry allows the downloading of the total power data streams for all
        the detectors, so that the sky-load differecing will be performed ``off-line''
        during ground data analysis.
        
\section{Offset balancing and 1/$f$ noise minimisation}
\label{sec:offset_balancing}

        In differential radiometers it is desirable to maintain the offset between the
        sky and reference signals as small as possible in order to minimise
        the effect of gain instabilities (see Eq.~(\ref{eq:dicke})).
        An offset introduced by an internal reference load can be balanced before
        differencing either by a variable back-end gain stage with a feed-back scheme
        to maintain the power output as close as possible to zero, or by
        by multiplying in software one of the two signals by a
        so-called {\em gain modulation factor}.
        In both cases the differential radiometer output can be written as:
        \begin{eqnarray}
                \label{eq:p_out}
                p_{\rm out} &=& a G_{\rm tot}k_{\rm B}\beta\left [\Ttsky + \Tn -
                r\left( \Ttref + \Tn \right) \right],\nonumber\\
                \Ttsky &=& \Tsky / \Lsky + (1 - \Lsky^{-1})\Tphys,\\
                \Ttref &=& \Tref / \Lref + (1 - \Lref^{-1})\Tphys,\nonumber
        \end{eqnarray}
        where $\Lsky$ and $\Lref$ are the insertion losses of the front-end sky and
        reference
        load antennas, $\Tphys$ is the physical temperature of the receiver front-end,
        $a$ is the detector proportionality constant, $G_{\rm tot}$ the radiometer total
        gain,
        $k_{\rm B}$ the Boltzmann constant, $\beta$ the radiometer bandwidth and $r$ the
        gain modulation factor. From Eq.~(\ref{eq:p_out})
        it follows that $p_{\rm out} = 0$ for $r = r_0^*$ where
        \begin{equation}
                r_0^* = \frac{\Ttsky+\Tn}{\Ttref+\Tn}.
                \label{eq:r_star_0}
        \end{equation}
        
        In the next section we will show how the above condition of zero power output
        represents a very good approximation of a balanced radiometer and
        is very effective in reducing the radiometer susceptibility to gain fluctuations
        to very low levels.
        
        In the case of \Planck-LFI the radiometer offset has been balanced
        through a software scheme that uses the undifferenced total power
        data to calculate $r$, which is expected to be constant on timescales
        as long as several days. This scheme has been preferred to a
        hardware gain modulation because it simplifies the
        radiometer hardware and avoids potential systematic errors from the
        variable gain stage.
        
        \subsection{Analytical derivation of the balancing condition}
        \label{subsec:analytical_treatment}
        
                In this section we calculate the value of the gain modulation factor that
                cancels the radiometer susceptibility to front-end gain fluctuations and
                show that this value is very well approximated by the the value $r_0^*$
                in Eq.~(\ref{eq:r_star_0}).
                
                Before proceeding into the details of
                the calculation we briefly examine the expected magnitude of gain and noise
                temperature
                fluctuations in the total power noise streams; more details concerning the
                analytical
                treatment of LFI pseudo-correlation radiometers can be found in 
                \citet{Seiffert:2002}.
                
                \subsubsection{Noise fluctuations in total power streams}
                \label{subsubsec:noise_total_power}
                
                        Cryogenic HEMT amplifiers are known to have 1/$f$ fluctuations in gain and noise
                        temperature 
                        \citep{Pospieszalski:1989, Wollack:1995, Jarosik:1996}.
                        The level of
                        these fluctuations can vary considerably among amplifiers and depends on the
                        details of device fabrication, device size, circuit design, and other factors.
                        Because of
                        this, we adopt an empirical model for the fluctuations also described in
                        \citet{Seiffert:2002}.
                        
                        In particular we write the 1/$f$ spectrum of the gain fluctuations as
                        \begin{equation}
                                \frac{\Delta G(f)}{G} = \frac{C}{f^\alpha},
                                \label{eq:delta_G_over_G}
                        \end{equation}
                        where $0.5 \lesssim \alpha \lesssim 1$  and $C$ represents a constant
                        normalization factor.
                        Similarly, we can write the noise temperature fluctuations as
                        \begin{equation}
                                \frac{\Delta T_{\rm n}(f)}{T_{\rm n}} = \frac{A}{f^\alpha},
                                \label{eq:delta_Tn_over_Tn}
                        \end{equation}
                        where $A \sim C /(2 N_{\rm s})$ is the normalization constant for noise
                        temperature fluctuations.
                        In the case of the {\sc Planck}-LFI radiometers we use the estimates
                        $A\simeq 1.8\times 10^{-5}$
                        for the 30 and 44 GHz and $A\simeq 2.5\times 10^{-5}$ for the 70 and 100 GHz
                        radiometers.
                        
                \subsubsection{Derivation of the optimal gain modulation factor for 1/$f$
                        minimisation}
                \label{subsubsec:derivation_rst}
                
                        Following the approach described in \citet{Seiffert:2002} it is
                        possible to derive analytical
                        formulas for the knee frequency relative to 1/$f$ fluctuations of the
                        differenced data
                        streams induced by front-end gain and noise temperature fluctuations.
                        
                        If we assume that gain and noise temperature fluctuations are correlated then we 
                        obtain the following relationship for the knee frequency:
                        \begin{eqnarray}
                                \fk^{\rm corr} &=& \left\{\frac{C}{\Ttsky + \Tn}\sqrt{\frac{\beta}{2}}
                            \left[(\Ttsky +(1+A/C)\Tn)\right.\right. \nonumber \\
                                &-& \left.\left.r(\Ttref +(1+A/C)\Tn)\right]\right\}^{1/\alpha},
                                \label{eq:fk_G+Tn}
                        \end{eqnarray}
                        which is zero for the following value of $r$:
                        \begin{equation}
                                r = r^*_{\rm corr} = \frac{\Ttsky+(1+A/C)\Tn}{\Ttref+(1+A/C)\Tn}.
                                \label{eq:r_star}
                        \end{equation}
                        
                        If we relax the hypothesis of completely correlated gain and noise temperature
                        fluctuations then we find:
                        \begin{eqnarray}
                          \fk^{\rm uncorr} &=& \left\{\frac{C}{\Ttsky + \Tn}\sqrt{\frac{\beta}{2}}
                          \left[\left((\Ttsky +\Tn)-r (\Ttref +\Tn)\right)^2\right.\right. 
                            \nonumber \\
                            &+& \left.\left. (1-r)^2 \Tn^2(A/C)^2\right]^{1/2}\right\}^{1/\alpha},
                          \label{eq:fk_G+Tn_uncorr}
                        \end{eqnarray}
                        which is minimised (but not nulled) by the following value of $r$:
                        \begin{equation}
                          \label{eq:r_star_uncorr}
                          r = r^*_{\rm uncorr} = \frac{(\Ttsky+\Tn)(\Ttref+\Tn) + (\Tn A/C)^2}
                          {(\Ttref+\Tn)^2 + (\Tn A/C)^2}.
                        \end{equation}
                        
                        Note that in both Eq.~(\ref{eq:r_star}) and (\ref{eq:r_star_uncorr}) the optimal 
                        value of $r$ is dependent on the ratio
                        $C/A \approx 2\sqrt{\Ns}$ where $\Ns$ is the number of amplifier stages; if $C/A
                        \gg 1$ then we have that in both cases the optimal value is approximated
                        by $r^*_0$ (given by Eq.~(\ref{eq:r_star_0})). 
                        
\begin{figure}[here]
                          \begin{center}
                            \resizebox{8.5 cm}{!}{\includegraphics{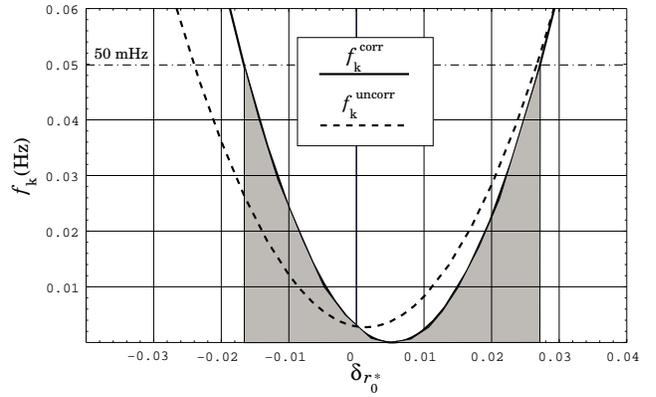}}
                          \end{center}
                          \caption{
                            Radiometer knee frequency versus the relative accuracy $\delta r/r = (r
                            - r^*_0)/r^*_0$
                            for the 70~GHz \Planck-LFI channel. The solid and dashed curves represent 
                            the
                            values of the knee frequency calculated assuming correlated and uncorrelated 
                            fluctuations
                            in the front-end amplifiers. The dashed-dotted 50 mHz line indicates the 
                            \Planck-LFI
                            design value for the knee frequency, while the shaded area represents the 
                            range of 
                            accuracy that is compliant with this design value.
                          }
                          \label{fig:fk_vs_delta_r}
                        \end{figure}

                        In Fig.~\ref{fig:fk_vs_delta_r} we show a plot of the knee
                        frequency calculated using Eq.~(\ref{eq:fk_G+Tn}) and (\ref{eq:fk_G+Tn_uncorr})
                        versus the relative accuracy $\delta_{r_0^*} = (r-r_0^*)/r_0^*$ for parameters 
                        typical
                        of the 70~GHz radiometer chain (which drives the accuracy requirements in the 
                        calculation of
                        $r$). 
                                
                        Two main consideration may be drawn from the results in 
                        Fig.~\ref{fig:fk_vs_delta_r}: first,
                        if $r = r^*_0$ then the radiometer knee frequency
                        is not nulled also if we assume complete correlation between gain and noise temperature 
                        fluctuations. 
                        In fact it can be shown \citep{Seiffert:2002} that gain fluctuations 
                        are cancelled
                        at first order and the residual 1/$f$ noise comes from noise temperature 
                        instabilities. Second,
                        if $r_0^*$ is known with an accuracy of the order of $\sim 1.5 \%$ then the 
                        resulting knee frequency
                        is below the design value (50~mHz in this case) independently from the hypothesis of correlated or 
                        uncorrelated fluctuations
                        in the front-end amplifiers.
                        
\begin{figure}[here]
                                \begin{center}
                                        \resizebox{9. cm}{!}{\includegraphics{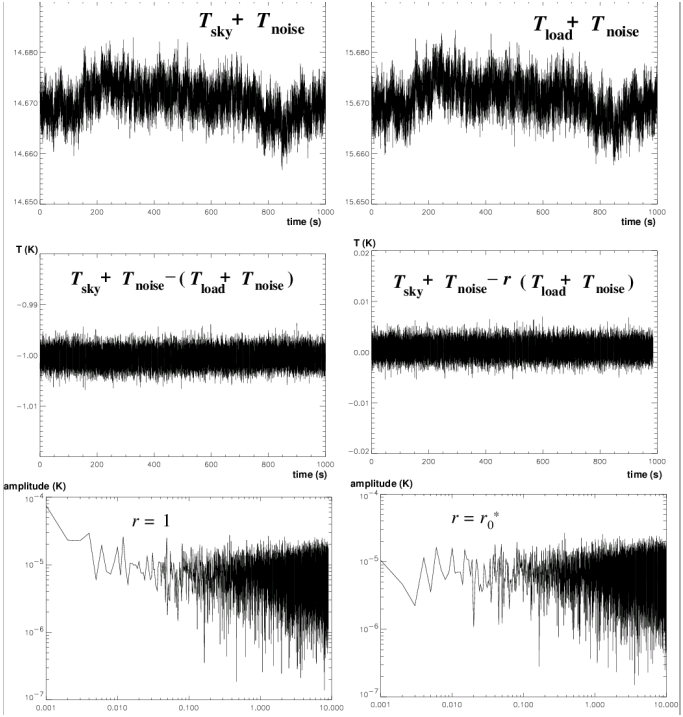}}
                                \end{center}
                                \caption{
                                        Simulated noise streams with parameters representative of a
                                        a 30 GHz LFI radiometer. The top graphs show the total power data (sky, left
                                        panel, and
                                        reference load, right panel), the middle graphs the differenced noise with $r=1$
                                        (left panel)
                                        and $r=r_0^*$ and the bottom graphs show the corresponding amplitude spectra.
                                }
                                \label{fig:noise_streams}
                        \end{figure}

                        In Fig.~\ref{fig:noise_streams} we show a simulation of the total power and of
                        the differenced
                        noise stream for a 30~GHz LFI radiometer in the hypothesis of correlated
                        gain and noise temperature fluctuations. The simulation has been performed
                        considering typical values for system noise temperature and front-end losses,
                        $\sim$1~K contribution from the telescope
                        (emissivity of 1\% for each reflector at 50~K physical temperature),
                        a physical temperature of 4.8~K for the reference load and a physical
                        temperature
                        of 20~K for the LFI focal plane.
                        
                        In the upper graphs of Fig~\ref{fig:noise_streams} we show the sky and reference
                        total power
                        data streams, while in the middle graphs we compare the differenced noise
                        streams
                        obtained with $r=1$ (left panel) and $r=r_0^*$ (right panel); the graphs show
                        that in both cases
                        most of the $1/f$ instability is removed, although in the $r=1$ case the 
                        sky-reference offset
                        is retained. A comparison between the two amplitude spectra (lower panels)
                        reveals that if the
                        ideal gain modulation factor is applied then the final knee frequency is much
                        lower compared to the
                        $r=1$ case\footnote{Note that in the case $r=r_0^*$ the knee frequency could not
                        be resolved due
                        to data stream length (1000 s) that limits the frequency resolution to $\sim
                        5$~mHz}.
                        
                        In Table~\ref{tab:r_values} we provide representative
                        estimates of $r^*_0$ for all the frequency channels of \Planck-LFI radiometers.
                        
                        \begin{table}[here]
                                \begin{center}
                                        \begin{tabular}{|c|c|c|}
                                                \hline
                                                30 GHz        & 44 GHz            &   70 GHz     \\
                                                \hline
                                                0.936         & 0.953          &  0.971      \\
                                                \hline
                                        \end{tabular}
                                        \caption{
                                                Estimates of $r^*_0$ for \Planck-LFI radiometers.
                                        }
                                        \label{tab:r_values}
                                \end{center}
                        \end{table}
                        
        \subsection{Susceptibility to other systematic effects}
        \label{subsec:expected_other_systematics}
        
                Apart from 1/$f$ noise, other instrumental systematic effects can be mitigated
                by balancing the offset in the radiometer output. In \Planck-LFI, for example,
                systematic effects can be expected from thermal variations of the 20~K and 300~K
                stages
                and from input bias fluctuations of the front-end amplifiers.
                
                Let us consider, for example, the
                effect caused by thermal instabilities in the front-end and in the back-end
                stages
                of the \Planck-LFI radiometers.
                Variations in the physical temperature in the radiometer couple with the
                measured output
                essentially through the following mechanisms:
                \begin{itemize}
                        \item variations in thermal noise induced by resistive front-end passive
                                components (feed-horn,
                                OMT and first hybrid);
                        \item amplifier gain oscillations (in the front-end and in the back-end);
                        \item amplifier noise temperature oscillations (in the front-end\footnote{
                                The $\sim$35 dB gain in the front-end makes negligible the effect from
                                fluctuations in the noise
                                temperature of the back-end amplifiers}).
                \end{itemize}
                The radiometer susceptibility can be written
                in terms of a transfer function, $\phi$, that links a physical temperature
                variation,
                $\delta \Tphys$, to the corresponding variation in the measured output, $\delta
                T_{\rm out}$,
                i.e.: $\delta T_{\rm out} = \phi \times\delta \Tphys$. 
                This function has been calculated following the approach discussed in
                \citet{Seiffert:2002} assuming complete correlation between the 
                temperature fluctuations at the level of the various radiometer components. This
                assumption is a reasonable one in the context of the \Planck-LFI radiometers where
                physical temperature variations induced in the
                20~K and 300~K stages will be dominated by relatively ``slow'' components
                with a very small contribution from harmonics with a frequency higher than $\sim$5~mHz.
                
                Under these assumptions the susceptibility to
                temperature variations
                in the 20 K stage ($\phi_{\rm FE}$) and in the 300 K back-end stage ($\phi_{\rm
                  BE}$) can be written as:
                \begin{eqnarray}
                        \phi_{\rm FE}&=& \Lsky\left\{(1-\Lsky^{-1}) - r (1-\Lref^{-1})+\right.
                    \nonumber\\
                        &+&\frac{\ln(10)}{10}\left[\Ttsky+\Tnfe-r(\Ttref+\Tnfe)\right]\frac{\partial
                        \Gfe^{\rm dB}}{\partial \Tphys}+
                    \nonumber\\
                        &+&\left.(1-r)\frac{\partial \Tnfe}{\Tphys}\right\},
                        \label{eq:Tf_FE_therm}
                \end{eqnarray}
                \begin{eqnarray}
                        \phi_{\rm BE} &=& \Lsky\frac{\ln(10)}{10}\left(\frac{\partial a^{\rm
                        dB}}{\partial \Tphys}
                        +\frac{\partial \Gbe^{\rm dB}}{\partial \Tphys}\right)\times\nonumber\\
                        &\times&\left[\Ttsky + \Tsys - r(\Ttref+\Tsys)\right],
                        \label{eq:Tf_BE_therm}
                \end{eqnarray}
                where $\Gfe^{\rm dB}$, and $\Tnfe$ are the gain in dB and the noise temperature
                of the front-end amplifiers, $\Gbe^{\rm dB}$
                the gain of the back-end amplifiers, $a^{\rm dB}$ the diode constant and
                $\Tphys$ the physical temperature.
                
                Note from Eq.~(\ref{eq:Tf_BE_therm})
                that the susceptibility to back end temperature fluctuations is cancelled if the
                radiometer
                output is balanced (i.e.
                $\phi_{f_{\rm BE}}=0$ for $r=r_0^*$); this is not only true for temperature
                variations, but
                for also for any instrumental effect causing instability in the back-end
                amplifiers.
                
                The susceptibility to front-end
                temperature variations, instead, is cancelled for a value of $r$ which is
                slightly different from $r_0^*$ (its expression
                is not reported here, but it can be derived from $\phi_{\rm FE}=0$ in
                Eq.~(\ref{eq:Tf_FE_therm})).
                In general the effect of any given systematic effect in the radiometer can be
                cancelled at first order using a particular value of the gain modulation factor;
                in practice,
                however, the choice of the ``best'' value of $r$ to use in the difference is
                driven by the
                most critical expected effect and by the feasibility to calculate $r$ from
                the measured data with the required accuracy.
                
                In \Planck-LFI, for example, with the condition $r=r^*_0$
                it is possible to minimise
                1/$f$ amplifier noise and effects from the back-end while keeping the
                radiometric
                susceptibility to other systematic effects at a
                level which allows to define realistic requirements on the thermal and
                electrical stability at the interfaces between the satellite and the instrument.
        
        \subsection{Required accuracy in the calculation of $r^*_0$}
        \label{subsec:required_accuracy}
        
                The main goal for balancing the radiometer output is to minimise the residual 1/$f$
                noise;
                therefore it is natural that the requirement
                on the gain modulation factor accuracy must be derived from the maximum knee
                frequency allowed
                in the experiment.
                
                From the curves shown in Fig.~\ref{fig:fk_vs_delta_r} we have derived a requirement
                on the accuracy of the
                gain modulation factor of the order of $\sim \pm$1.5\%.
                
                In some cases, however, the accuracy needed in the calculation of the gain modulation
                factor may
                be determined by the level of other effects; in \Planck-LFI radiometers, for
                example, the value $r_0^*$ will not only minimise 1/$f$ noise, but also effects from
                back-end thermal instabilities;
                in this case the need for a very low radiometer susceptibility to back-end
                temperature fluctuations imposes a tight requirement on the accuracy of the gain modulation
                factor, ranging from $\sim\pm 1\%$ at 30~GHz to $\sim \pm 0.3\%$ at 70~GHz.
        
\section{Calculation of $r^*_0$ from measured data}
\label{sec:calculation_of_r}

        During a long-duration CMB measurement the radiometer noise
        properties are subject to slight, slow variations in time due to
        thermal/electrical drifts and ageing of the radiometer electronic
        components. If the offset balancing is done in software, then the
        value of $r^*_0$ can be recalculated and updated (if necessary)
        in the data reduction software to compensate for such changes.
        
        As we show in detail in this section,
        the gain modulation factor can be calculated from radiometer data acquired in
        ``total power mode'', i.e. before differencing is performed. 
        Although the current telemetry bandwidth available to the LFI instrument
        allows to download the complete total power data streams, 
        it is possible to calculate $r$ from a limited portion (about $\sim$15 min) of these data. 
        This implies that it could be possible to perform the sky-load differencing
        on-board (with a saving of $\sim$50\% in telemetry bandwidth) providing margin
        for possible non-nominal operation scenarios.

        In this section we discuss the following three calculation methods and compare them 
        from the point of view of the obtainable accuracy:
        \begin{enumerate}
                \item $r^*_0$ calculated from the ratio of the average sky and reference load
                        levels;
                \item $r^*_0$ calculated from the ratio of the sky and reference load standard
                        deviations;
                \item $r^*_0$ calculated by minimising of the final differenced data stream knee
                        frequency.
        \end{enumerate}
        In the following discussion we will consider only the presence of white+1/$f$
        noise in the
        radiometer data streams and assume the availability of 15 min of total power
        data
        (as in the case of \Planck-LFI) in order to evaluate the efficiency of each
        method. The effect of the presence
        of the astrophysical signal and of other instrumental systematic effects is
        discussed in
        Sect.~\ref{sec:impact_of_systematic_effects}.
        
        \subsection{Calculation from average signal level}
        \label{subsec:calc_from_signal_level}
        
                The simplest way to calculate $r$ from the total power data is to use directly
                the definition
                provided by Eq.~(\ref{eq:r_star_0}), i.e. to take the ratio between the average
                signals
                acquired when looking at the sky,
                $\Vskyi = G_i(\Ttsky + \Tni)$,
                and when looking at the reference load, $\Vrefi = G_i(\Ttref + \Tni)$. In this
                case $r$ is given by:
                \begin{equation}
                r = \frac{\sum_{i=1}^{N}\Vskyi}{\sum_{i=1}^{N}\Vrefi},
                \label{eq:r_from_signal}
                \end{equation}
                where $N$ represents the number of samples available in the data stream.
                Now we know that although gain and the noise temperature will display 1/$f$
                fluctuations, thanks to the pseudo-correlation scheme
                they will be the same (at first order) in both the sky and reference signals.
                Therefore for each sample, $i$, we can write: $G_i = G + \delta G_i^{1/f}$
                and $\Tni = \Tn + \delta \Tni^{1/f}$, where $G$ and $\Tn$ represent
                the average gain and noise temperature level. Furthermore there will be
                a white nose component $\deltaTWN$ such that
                $\langle \deltaTWN \rangle = 0$ and $\sigmaWN = (\Ttskyref +
                \Tn)/\sqrt{\beta\tau}$.
                
                Let us now evaluate the impact of gain and noise temperature fluctuations
                separately.
                If the total power noise streams contain 1/$f$ gain fluctuations then we can
                write the sky and load data streams as:
                \begin{equation}
                \Vskyrefi =(G+\delta G_i^{1/f})(\Ttskyref + \Tn ) + \deltaTWN;
                \end{equation}
                therefore the sums in Eq.~(\ref{eq:r_from_signal}) have the form:
                \begin{eqnarray}
                \label{eq:sum_vskyrefi_dG}
                \sum_{i=1}^{N}\Vskyrefi &=& G N \left( \Ttskyref + \Tn \right)\times\\
                &\times&\left[1+\frac{1}{N}\sum_{i=1}^{N}\frac{\delta G_i^{1/f}}{G}\right]
                +\sum_{i=1}^{N}\deltaTWN.\nonumber
                \end{eqnarray}
                
                Similarly, in the case of 1/$f$ noise temperature fluctuations then
                Eq.~(\ref{eq:sum_vskyrefi_dG}) becomes:
                \begin{eqnarray}
                \label{eq:sum_vskyrefi_dTn}
                \sum_{i=1}^{N}\Vskyrefi &=& G N \left( \Ttskyref + \Tn \right)\times\nonumber\\
                &\times&\left[1+N^{-1}\sum_{i=1}^{N}\frac{\delta T_{{\rm n,}i}^{1/f}}{\Ttskyref
                + \Tn }\right]
                +\\
                &+&\sum_{i=1}^{N}\deltaTWN.\nonumber
                \end{eqnarray}
                
                In both Eqs.~(\ref{eq:sum_vskyrefi_dG}) and (\ref{eq:sum_vskyrefi_dTn}) the two
                sums
                on the right-hand side tend to zero for large values of $N$, so that the ratio
                of the average signals $\rightarrow r_0^*$ for $N\rightarrow \infty$, which
                implies that
                Eq.~(\ref{eq:r_from_signal}) can be used also in
                presence of 1/$f$ noise provided that the statistics is sufficiently large.
                
\begin{figure}[here]
                        \begin{center}
                                \resizebox{8. cm}{!}{\includegraphics{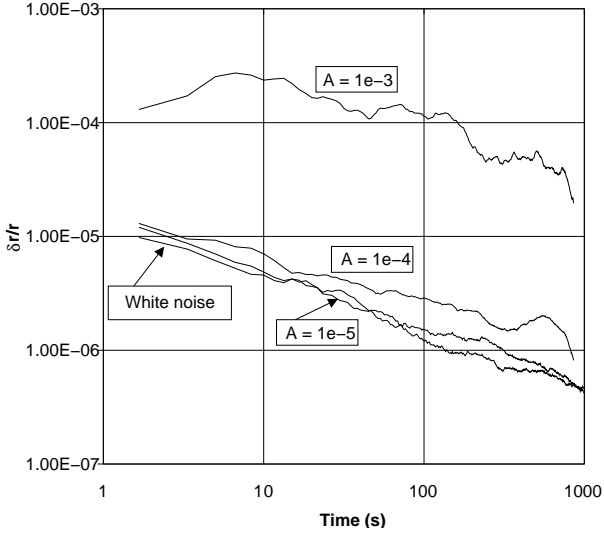}}
                        \end{center}
                        \caption{
                                Effect of gain and noise temperature 1/$f$ fluctuations on the
                                calculation
                                of $r$ from the ratio of the average signal levels. Each curve is an average of
                                50 noise
                                realisations relative to the 30 GHz \Planck-LFI channel.
                        }
                        \label{fig:conv_from_signal}
                \end{figure}

                In Fig.~\ref{fig:conv_from_signal} we show the result of a numerical simulation
                of the convergence of
                Eq.~(\ref{eq:r_from_signal}) to the theoretical value $r_0^*$.
                Each curve represents the value of $\delta r / r_0^*$ versus time, averaged over
                50 equivalent
                noise realisations of the 30~GHz \Planck-LFI channel. The different curves are
                relative
                to different levels of 1/$f$ noise in both gain and noise temperature
                ($A = 0$, i.e. white noise only, $A = 10^{-5}$, $A = 10^{-4}$, $A = 10^{-3}$
                i.e.
                low, intermediate and high level of 1/$f$ noise).
                The figure shows that the presence of 1/$f$ fluctuations determines a slower
                convergence
                with respect to the white noise case. Even in the worst case, however,
                (which represents an unrealistic case with respect to the expected LFI
                performances) the accuracy
                obtained with 15 minutes of data is much better than 0.01\%.
                
                In order to calculate properly the ratio $r$ from Eq~(\ref{eq:r_from_signal}),
                it is necessary that any ``zero-level error'' introduced
                by the back-end electronics is small. Let us evaluate the effect of a constant
                offset $\delta V$
                added to both states in the switching output of the radiometer; the voltage
                measured in the two switch
                states can be written as:
                \begin{equation}
                        \Vskyref = G(\Ttskyref + \Tn)+\delta V .
                \end{equation}
                
                If the offset is unknown, it introduces an uncertainty given by:
                \begin{equation}
                        \delta r = \frac{\Vsky}{\Vref}-\frac{\Vsky+\delta V}{\Vref + \delta V} =
                        \frac{\Vsky-\Vref}{\Vref(\Vref+\delta V)}\delta V.
                        \label{eq:error_from_offset}
                \end{equation}
                
                From Eq.~(\ref{eq:error_from_offset}) it is straightforward to obtain the
                maximum values of the
                unknown offsets that would produce an uncertainty on $r$ equal to the required
                accuracy. In the
                case of \Planck-LFI these values (in temperature units) range from $\sim\pm 0.5$
                K at 30 GHz to
                $\sim\pm 2.34$ K at 100 GHz, which are compatible with the current electronics
                design.

                Finally we underline that the above treatment is valid in the hypothesis of 
                detector linearity and negligible gain compression. In general, however, these
                conditions are largely met in CMB anisotropy experiments, where the signals to 
                be measured are small. In the case of the \Planck-LFI radiometers, for example,
                the radiometer output power is of the order of $-25$~dBm, which is much smaller than
                the $-1$-dB compression point of the HEMT amplifiers that is of the order of 
                $\sim -15$~dBm and falls well in between the range of detector linearity 
                ($\sim -35 \div -15$~dBm).
                
        \subsection{Calculation from standard deviations}
        \label{subsec:calc_from_sigma}
                A second approach is to calculate $r$ from the ratio of the standard deviations
                of the total power noise streams, i.e.:
                \begin{equation}
                        r = \frac{\sigmaskyN}{\sigmarefN} = \left[
                        \frac{\sum_{i=1}^{N}\left(\Vskyi-\langle \Vsky \rangle\right)^2}
                        {\sum_{i=1}^{N}\left(\Vrefi-\langle \Vref \rangle\right)^2}\right]^{1/2}.
                        \label{eq:r_from_stdev}
                \end{equation}
                
                An advantage of this method compared to the previous one is that the
                ratio of standard deviations is insensitive to spurious offsets introduced by
                the back-end
                electronics; let us evaluate now the accuracy that is obtainable in presence of
                noise.
                In the case the data streams are characterised by white noise only then we have
                that $\sigmaskyrefN \rightarrow (\Ttskyref + \Tn)/\sqrt{\beta\tau}$ for
                $N\rightarrow\infty$,
                so that for large values of $N$ Eq.~(\ref{eq:r_from_stdev}) approximates
                $r_0^*$.
                
                Of couse, it is also well known that the standard deviation
                of an infinite time stream of 1/$f$-type noise formally diverges.
                For the case under consideration here, the ratio of standard
                deviations of finite sets of white
                noise plus some 1/$f$-type noise, 
                we examine below the details of
                accuracy and convergence. 
                
                Let us consider first the effect of gain fluctuations; from
                Eq.~(\ref{eq:sum_vskyrefi_dG}) we have that
                \begin{eqnarray}
                        \langle \Vskyref \rangle_N &=& \frac{1}{N}\sum_{i=1}^{N}\Vskyrefi = \nonumber \\
                        &=& G \left(\Ttskyref + \Tn\right)\left(1+\left\langle\frac{\delta
                        G^{1/f}}{G}\right\rangle_N\right)+\nonumber\\
                        &+&\langle \deltaTWN \rangle_N,
                \end{eqnarray}
                and, consequently,
                \begin{eqnarray}
                        \label{eq:vskyi_vsky_dg}
                        &&\sum_{i=1}^N\left(\Vskyrefi-\langle \Vskyref \rangle \right)^2 \approx
                        \nonumber\\
                        &\approx & \left(\Ttskyref +\Tn\right)^2\sum_{i=1}^N \left(\delta G_i^{1/f}-
                        \langle \delta G^{1/f} \rangle \right)^2 +\nonumber\\
                        &+&\sum_{i=1}^N\left(\deltaTWNi - \langle\deltaTWN\rangle\right)^2
                        \stackrel{N\rightarrow\infty}{\longrightarrow} \\
                        &\stackrel{N\rightarrow\infty}{\longrightarrow}& (N-1)
                        \left[\left(\Ttskyref+\Tn\right)^2\sigma^2_{G^{1/f}}+
                        \left(\sigmaWN\right)^2\right]\nonumber .
                \end{eqnarray}
                
                Considering that $\sigmaWN \stackrel{N\rightarrow\infty}\longrightarrow
                (\Ttskyref + \Tn)/\sqrt{\beta\tau}$ then it follows that
                \begin{equation}
                        \sigmaskyrefN \stackrel{N\rightarrow\infty}{\longrightarrow} \left(\Ttskyref +
                        \Tn\right)
                        \left(\sigma_{G^{1/f}}+\frac{1}{\sqrt{\beta\tau}}\right),
                \end{equation}
                so that the ratio of the total power standard deviations converges to $r_0^*$
                for large $N$.
                
                Let us now consider the presence of 1/$f$ fluctuations in the noise temperature.
                In this case Eq.~(\ref{eq:vskyi_vsky_dg}) has the form:
                \begin{eqnarray}
                        &&\sum_{i=1}^N\left(\Vskyrefi-\langle \Vskyref \rangle \right)^2 \approx
                        \nonumber\\
                        &\approx & G^2\sum_{i=1}^N \left(\delta T_{{\rm n},i}^{1/f}-
                        \langle \delta T_{\rm n}^{1/f} \rangle \right)^2 + \nonumber\\
                        &+&\sum_{i=1}^N\left(\deltaTWNi - \langle\deltaTWN\rangle\right)^2
                        \stackrel{N\rightarrow\infty}{\longrightarrow} \\
                        &\stackrel{N\rightarrow\infty}{\longrightarrow}& (N-1)\left[G^2\sigma^2_{T_{\rm
                        n}^{1/f}}+
                        \left(\sigmaWN\right)^2\right]\nonumber .
                        \label{eq:vskyi_vsky_dTn}
                \end{eqnarray}
                
                It follows that the ratio of standard deviations converges, in this case, to:
                \begin{equation}
                        \sqrt{\frac{G^2\sigma^2_{T_{\rm n}^{1/f}}+(\sigmaskyWN)^2}
                        {G^2\sigma^2_{T_{\rm n}^{1/f}}+(\sigmarefWN)^2}} = r_0^*
                        \sqrt{\frac{\left(G\sigma_{T_{\rm n}^{1/f}}/\sigmaskyWN\right)^2+1}
                        {\left(G\sigma_{T_{\rm n}^{1/f}}/\sigmarefWN\right)^2+1}} \neq r_0^*.
                \end{equation}
                
                This clearly shows that in presence of noise temperature 1/$f$ fluctuations the
                ratio of standard
                deviations does not converge to $r_0^*$.
                In particular with increasing amplitude of the 1/$f$ noise temperature
                fluctuations
                (i.e. for increasing values of $A$) we have that $G\sigma_{\Tn}^{1/f} \gg
                \sigmaWN$
                and the ratio of standard deviations
                tends to 1. We can therefore calculate the following lower limit for the
                relative accuracy
                on $r$ obtained by this method:
                \begin{equation}
                        \frac{\delta r}{r}\stackrel{A\rightarrow\infty}{\longrightarrow}\frac{1-
                        r_0^*}{r_0^*}.
                \end{equation}
                
\begin{figure}[here]
                        \begin{center}
                                \resizebox{8. cm}{!}{\includegraphics{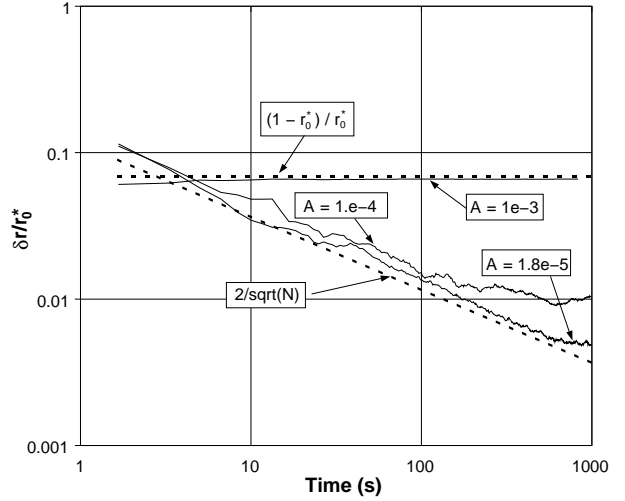}}
                        \end{center}
                        \caption{
                                Effect of noise temperature 1/$f$ fluctuations on the calculation of
                                $r$ from the ratio
                                of standard deviations signal levels.
                                Each curve is an average of 50 noise
                                realisations relative to the 30 GHz \Planck-LFI channel.
                        }
                        \label{fig:conv_from_rms}
                \end{figure}

                In Fig.~\ref{fig:conv_from_rms} we show the results of numerical simulations
                relative to the
                30~GHz \Planck-LFI channel showing the
                convergence of the ratio of noise standard deviations
                (Eq.~\ref{eq:r_from_stdev})
                for various levels of 1/$f$ noise.
                The two dashed curves represent the limits for $A\rightarrow 0$ (i.e. white
                noise only) and
                $A\rightarrow\infty$. In the first case (white noise), because the standard
                deviation of $N$ random samples
                tends to the standard deviation of the gaussian distribution as $\delta\sigma /
                \sigma = 1/\sqrt{N}$,
                we have that the accuracy calculated by the ratio
                of the standard deviations increases with the number of samples as
                $\delta r/r = 2/\sqrt{N}$. In the second case we have that the ratio of standard
                deviations
                does not converge to $r_0^*$ but to $\frac{1-r_0^*}{r_0^*}$. The figure clearly
                shows that for increasing values of $A$ the behaviour changes between these two
                limits.
                
                When the 1/$f$ noise level is high it is still possible to estimate
                $r$ from the ratio of the
                white noise rms that can be derived either by fitting the high frequency end of the
                noise power spectrum or using a proxy of the white noise level by means of the
                two-point variance function $\sigma^2(i-j)=\langle (x_i - x_j)^2 \rangle$
                as described by \citet{Janssen:1996}. In this case $r$ will converge
                to $r_0^*$ with an upper limit accuracy of $\sim 2/\sqrt{N}$
                (see Fig. \ref{fig:conv_from_rms}).
                
        \subsection{Calculation by minimising the final knee frequency}
        \label{subsec:calc_minimsing_fk}

                A third approach that we consider here is the optimisation of the gain
                modulation
                factor by defining a window $[r_{\rm min},r_{\rm max}]$
                around $r_0^*$ and then finding
                the ``best'' value of $r$ that minimises knee frequency of the differenced noise
                stream.
                
                Although this method probes directly the 1/$f$ noise characteristics of the
                final differenced data,
                there are some limitations and caveats that are worth mentioning:
                
                \begin{enumerate}
                        \item {\bf frequency resolution and accuracy}.
                                If $\Delta t$ represents the length (in time) of the data available for
                                analysis, then the absolute minimum frequency that can be resolved in Fourier space     
                                is $f_{\rm min}\sim 2/\Delta t$. For $\Delta t \sim$ 900~s (15 minutes) we have that 
                                $f_{\rm min}\sim$ 2~mHz.
                                Although the data length can be enough to resolve
                                frequencies $\sim 1$ mHz, in order to determine $f_{\rm k}$ we must be able to 
                                recognise the
                                frequency at which the 1/$f$ noise has an amplitude equal to the rms value of the 
                                white noise.
                                Because noise dominates, each individual sample has 100\% uncertainty so that we
                                must rebin the data with a consequent loss in frequency resolution.
                
                        \item {\bf Presence of astrophysical signal in differenced noise stream}.
                                When the sky and load data streams are differenced taking into account a value
                                of $r$
                                in the interval $[r_{\rm min},r_{\rm max}]$, the differenced noise stream will
                                also contain
                                the astrophysical signal (dipole, galaxy, etc.) at all the harmonics of the
                                scan frequency ($\sim$16~mHz for \Planck). This implies that with the real data
                                the
                                calculation of the knee frequency will be affected, in principle, by the power
                                at the first
                                10 harmonics or so of the scan frequency. Therefore some assumptions on the
                                expected signal
                                are probably necessary to remove it from the differenced data before computing
                                $f_{\rm k}$.
                
                \end{enumerate}
        
\section{Application to laboratory radiometer data}
\label{sec:example_experimental}

        The gain modulation strategy described in the previous sections is routinely
        applied with success in the \Planck-LFI prototype radiometers. Here we present
        some results obtained with experimental data from the 30~GHz LFI {\em Elegant BreadBoard} (EBB)
        radiometer  assembled at the Jodrell Bank Observatory (JBO) in the framework of the \Planck\
        collaboration. Further details about the EBB testing campaign of the \Planck-LFI 
        radiometers will be discussed in a dedicated set of papers which is currently in preparation.
        
        The EBB 30~GHz radiometer is comprised of a front-end module (FEM) manufactured at JBO and 
        a back-end module (BEM) manufactured at the University of Cantabria (Santander, Spain) connected by
        waveguides approximately 0.75~m long. Each FEM channel includes a HEMT phase switch which has also
        been designed and manufactured by JBO. In the tests described here the FEM was operated at
        20~K while the BEM was maintained at room temperature.

        Because the aim of these tests was to verify the validity of the differential concept rather
        than to achieve flight-like performances, the test setup was kept as simple as possible; in particular
        the front-end phase switches were operated at 280~Hz (instead of 4096~Hz as in the LFI baseline)
        and the 300~K thermal environment was not maintained at a high degree of stability.
        
        Before the stability tests were carried out, the BEM white noise floor
        was measured to be significantly less than the noise level with the
        FEM and BEM connected together. Fifteen minute tests were then carried out with one
        of the phase switches kept in a fixed state. A 280~Hz square wave switch
        waveform for the other phase switch was generated by the data acquisition card
        counter outputs, under LabVIEW control.
        The detected signals were therefore 280 Hz square waves, which were sampled at
        504000 samples per second. After discarding 20\% of the data (10\% from each edge
        of the square wave), each switched state  was averaged to obtain a single
        number, A$_1$, B$_1$, A$_2$ or B$_2$, where A represents the `on' state for each
        channel and B the `off' state.
        
\begin{figure}[here]
          \begin{center}
            \resizebox{8.5 cm}{!}{\includegraphics{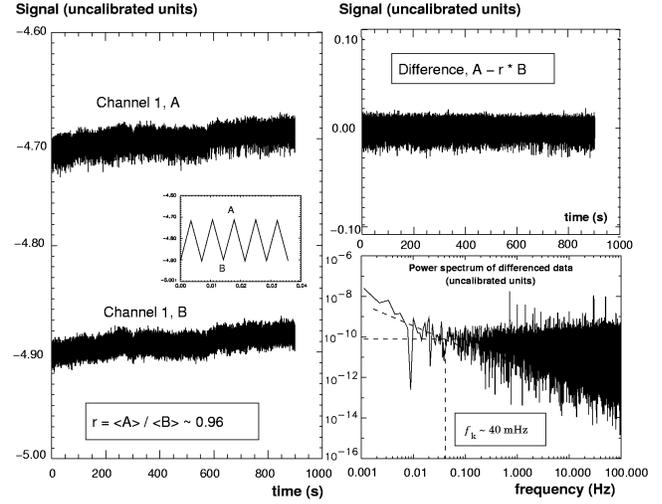}}
          \end{center}
          \caption{
            Measured data from the 30~GHz \Planck-LFI prototype radiometer. The
            graph on the left shows about 15~min of total power noise data from one of the two 
            radiometer
            channels. The small inset shows how measurements actually appear at the output of 
            each detector, i.e. as an alternating series of A and B values. The two panels
            on the right show the differenced noise stream in time and frequency domain. The
            final knee frequency is $\sim$40~mHz.
          }
          \label{fig:sky_load_exp}
        \end{figure}

        In the left graph of Fig.~\ref{fig:sky_load_exp} we show the total power levels
        of one of the two output radiometer detectors, the top trace representing the A states
        and the lower the B states. The small inset shows how the actual averaged
        diode output looks, i.e. a series of A and B
        values alternating at half the frequency of the phase switch;
        in the big graph, instead,
        the measured values relative to the two switch states are shown separately for
        better clarity. 
        
        From the ratio of the average levels of the two data streams we have calculated
        a value $r = 0.96075$ that has been applied to produce the differenced noise 
        stream shown in the top panel
        of the right part of Fig.~\ref{fig:sky_load_exp}. The graph shows that the offset has been 
        removed 
        as well as the long-timescale instabilities that are evident in the total power data. 
        The power spectrum of the differenced data shows a final knee
        frequency of the order of $\sim$ 40~mHz\footnote{the knee frequency 
          was defined as the frequency at which the power spectrum
          has a value equal to $\sqrt{2}$ of the average white noise level}.
        
        We also calculated $r$ by taking the ratio of the standard deviation of the data
        streams, and by directly minimising the final knee frequency. In the first case we
        obtained a value of $r = 0.95568$, 
        while in the second case we found $r=0.96076$ which is practically coincident
        with the one calculated with the ratio of signal levels.
        
\begin{figure}[here!]
          \begin{center}
            \resizebox{7. cm}{!}{\includegraphics{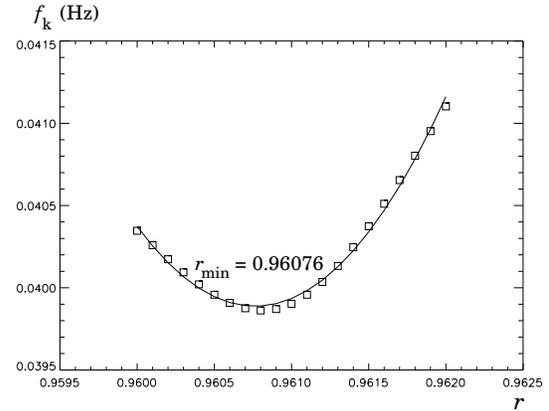}}
          \end{center}
          \caption{
            Behaviour of $\fk$ versus $r$ around the minimum. The curve was obtained by applying
            different values of $r$ to the total power data shown in Fig.~\ref{fig:sky_load_exp} 
            and 
            then calculating
            the knee frequency of the differenced data. The continuous line is a polynomial fit 
            of the discrete     data points.
          }
          \label{fig:rmin}
        \end{figure}

        In Fig~\ref{fig:rmin} we show the behaviour of the
        knee frequency versus $r$ around the minimum;
        the continuous line represents a polynomial fit used to determine the minimum.
        
        Although the knee frequency obtained from the differential data is within the \Planck-LFI requirements
        its value is higher than the theoretical value ($\lesssim 10$~mHz) that can be estimated from
        Eqs.~(\ref{eq:fk_G+Tn}) and (\ref{eq:fk_G+Tn_uncorr}). 
        
\begin{figure}[here!]
          \begin{center}
            \resizebox{8.5 cm}{!}{\includegraphics{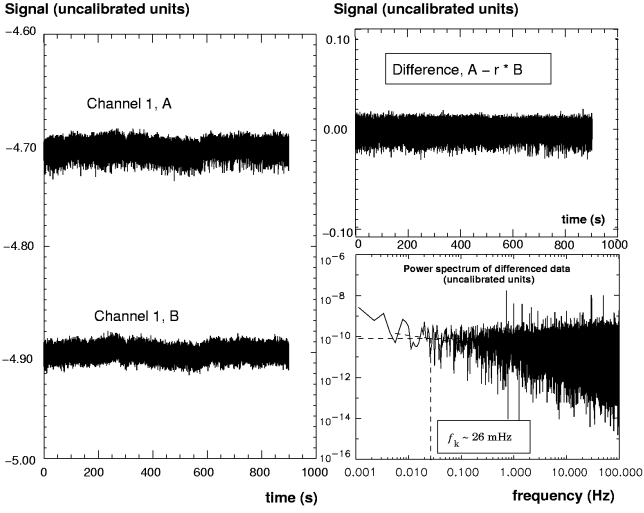}}
          \end{center}
          \caption{
            Left panel: total power radiometer data after the removal of the linear
            trend.
            Right panel: differetial output in time (top) and 
            frequency domains (power spectrum, bottom). The knee frequency was estimated 
            $\sim 26$ mHz.
          }
          \label{fig:data_notrend}
        \end{figure}

        Part of this discrepancy was caused
        by thermal drifts in the 300~K environment (that were observed to be of the order
        of 1.3 K/hr and can be clearly seen in the total power data streams);  to evaluate 
        this effect we 
        removed the linear trend from the data and then calculated the spectrum of the
        resulting differenced time stream. In Fig.~\ref{fig:data_notrend} we show the
        total power and the differential radiometric outputs after the removal of the
        linear trends; the noise power spectrum shows that the final knee frequency
        was reduced from $\sim 40$ to $\sim 26$ mHz.
        
        The remaining discrepancy of about $\sim$15~mHz 
        was likely to be caused by other non-idealities in the experimental setup which have not been
        completely understood yet.
        
        These results demonstrate how the pseudo-correlation scheme adopted for the LFI radiometers 
        is effective in reducing the effect of amplifier instability to very low levels also in
        experimental conditions far from the nominal flight operations.
        Furthermore, we have shown that the gain modulation factor can be determined with
        very simple analysis of a limited portion of the total power radiometer data.

\section{Impact of main systematic effects on the calculation of $r$}
\label{sec:impact_of_systematic_effects}

        In this section we present a brief analysis of the impact of some systematic
        effects on the calculation of the gain modulation factor.
        We can identify the following three possible classes of effects that can impact on the 
        calculation of $r$:
        
        \begin{enumerate}
        
                \item  effects that induce a variation of the total power signal levels, which
                        have an influence especially in the case $r$ is calculated
                        using the ratio of the average total power sky and load levels;
        
                \item  effects that induce a variation of the standard deviation of the total
                        power
                        noise streams, which have an effect in the case
                        $r$ is calculated using the ratio of standard deviations;
        
                \item  effects that change the low frequency spectral behaviour of the noise
                        streams and
                        limit the ability to derive $r$ from the analysis of the differenced noise
                        stream
                        frequency spectrum.
        
        \end{enumerate}
        
        Considering that the most accurate method to derive $r$ is from
        the ratio of the average signal levels, we focus on the first class of effects,
        which can
        be subdivided into the two following sub-classes:
        
        \begin{enumerate}
        
                \item effects that induce roughly the same variation in both the sky and
                        reference load
                        total power signals ({\em symmetric effects}), and
        
                \item  effects that introduce variations in only one of the two signals
                        (sky or reference load, {\em asymmetric effects}).
        
        \end{enumerate}
        
        Effects that belong to the first category are, for example, fluctuations induced
        by radiometer
        temperature instabilities and by bias voltage variations. Effects belonging to
        the second category
        are, for example, the CMB dipole, telescope temperature fluctuations, reference
        load variations.
        
        \subsection{Symmetric effects}
        \label{subsec:symmetric}
        
                Let us consider a systematic effect that causes a variation $\delta T$
                in both the sky and reference load total power signals. Clearly this is
                completely
                equivalent to the case of a constant offset introduced by the back-end
                electronics
                discussed in Sect.~\ref{subsec:calc_from_signal_level}. Therefore we can
                conclude that we need systematic effects with an amplitude of the order of 1~K
                (which
                is 2-3 order of magnitudes higher compared to what is expected for \Planck-LFI)
                or more to
                to introduce uncertainties in the value of $r$ greater than 1\%.
                
                
                
        \subsection{Asymmetric effects}
        \label{subsec:asymmetric}
        
                Let us now consider a systematic effect that causes a
                variation $(\delta T)^{\rm sky}$ only in the sky power signal. The relative
                uncertainty on the
                value of $r$ determined by this effect is:
                \begin{equation}
                        \frac{\delta r}{r_0^*} = \frac{(\delta T)^{\rm sky}}{\Ttsky + \Tn}.
                \end{equation}
                
                In the case we have a systematic variation only in the reference load signal
                then the uncertainty on $r$ is:
                \begin{equation}
                        \frac{\delta r}{r_0^*} = \frac{(\delta T)^{\rm ref}}{\Ttref + \Tn + (\delta
                        T)^{\rm ref}}
                \end{equation}
                
                In Table~\ref{tab:asymmetric_effects} we report, for the \Planck-LFI 30 and
                100~GHz
                channels, the values of the sky and reference load
                variations that would determine a change in $r$ equal to the needed accuracy
                (first two rows),
                together with estimates of the variation in the sky and reference load signals
                due to the major expected asymmetric effects (last three rows).
                
                \begin{table}[here]
                        \begin{center}
                                \begin{tabular}{|l|c|c|}
                                        \hline
                                                    &          30 GHz                &
                                        100 GHz     \\
                                        \cline{2-3}
                                                    & \multicolumn{2}{c|}{values in mK} \\
                                        \hline
                                        $(\delta T)_{\rm max}^{\rm sky}$    & $\pm 146$           &      $\pm$ 112    \\
                                        \hline
                                        $(\delta T)_{\rm max}^{\rm ref}$    & $\pm 154$           &      $\pm$ 113    \\
                                        \hline\hline
                                        $(\delta T)_{\rm dip.}$             & $\pm 3.4$           &      $\pm$ 2.7    \\
                                        \hline
                                        $(\delta T)_{\rm tel.}$             & $<\pm 1$            & $<\pm 1$
                                        \\
                                        \hline
                                        $(\delta T)_{\rm ref}$              & $<\pm 1$            & $<\pm 1$
                                        \\
                                        \hline
                                \end{tabular}
                                \caption{
                                        Impact of asymmetric effects on the calculation of the gain modulation
                                        factor
                                        for the \Planck-LFI radiometers. The first two rows show the variation in the
                                        sky and reference
                                        signal that would cause a change in $r$ equal to the needed accuracy of $\pm
                                        1\%$ at 30~GHz and
                                        of $\pm 0.2\%$ at 100~GHz. The last three
                                        rows list the expected sky and reference signal variations caused by the major
                                        sources
                                        of asymmetric effects (CMB dipole, telescope temperature fluctuations, reference
                                        load
                                        instabilities). \label{tab:asymmetric_effects}
                                }
                        \end{center}
                \end{table}
                
                The values reported in Table~\ref{tab:asymmetric_effects} show that in the
                \Planck-LFI case
                the calculation of the gain modulation
                factor from the ratio of the average signal level is largely immune from any
                expected
                systematic variations in the sky signal and/or in the reference load.
                This means that in this case $r$ can
                be considered constant at first order and the only significant changes are
                expected from long timescale
                variations of the radiometer noise properties. These changes will be recognised
                in the ground data analysis and compensated for by periodically updating
                the value of $r$ during data reduction.

\section{Conclusions}
\label{sec:conclusions}
        High sensitivity CMB radiometric measurements require a very low susceptibility
        of the receiver
        to amplifier 1/$f$ noise, which can be obtained only by differential
        measurements. The pseudo-correlation
        differential radiometer is a receiver particularly suitable for
        CMB radiometric measurements, allowing differential measurements without the
        need of an active front-end switch.
        
        In this scheme the best 1/$f$ noise suppression is reached when the
        radiometer is completely balanced. For slightly unbalanced configurations,
        e.g. when a cryogenic internal load is used as a reference signal, the
        effects introduced by the offset between the two inputs can be
        minimised after detection either by a variable
        DC gain stage that maintains the differenced output as close as possible to
        zero, or by
        multiplying in software one of the two signals by a factor, $r$, that must be
        properly tuned to
        balance the receiver. At first order the parameter $r$ is equal to the ratio
        between the sky and reference ``total power'' signal
        levels, i.e. $r\approx (\Ttsky + \Tn)/(\Ttref + \Tn)$. With this value it is
        possible
        to obtain differenced noise streams with a knee frequency of the order of few
        tens of mHz
        and make the radiometer insensitive to fluctuations in the back-end amplifier
        gain.
        
        The accuracy required in the calculation of $r$ from ground data must be
        determined
        by the instrument requirements on 1/$f$ noise and other systematic
        effects. In the case of \Planck-LFI the main driver is the susceptibility to
        thermal
        fluctuations in the warm back-end stage, which can be maintained to required
        levels
        if $r$ is calculated with an accuracy ranging from $\pm$1\% for the 30~GHz
        channel to $\pm$0.3\% for the 70~GHz channel.
        
        In our study we have considered three methods for calculating $r$, using total
        power
        radiometer data.
        
        
        The most straightforward scheme uses the ratio of the average level of the total
        power signals.
        This method proves quite simple, accurate and relatively immune from systematic
        effects like 1/$f$ amplifier fluctuations, thermal effects etc. In order to be
        applied it is necessary
        that any spurious offset introduced by the back-end electronics is known with a
        relative accuracy
        of the order of $\pm$4\%. This is the baseline method in the case of \Planck-LFI.
        
        The second method uses the ratio of the standard deviations of the total power
        noise streams
        and is limited by its sensitivity to 1/$f$ fluctuations of the front-end
        amplifier
        noise temperature, which sets a lower limit on the expected accuracy that is
        about one order of
        magnitude worse than method 1.
        
        The third method uses a minimisation strategy of the knee frequency of the
        differenced noise stream, by which it is possible to
        obtain the same level of accuracy of method 1, provided that the lenght of the total power data 
        stream is
        enough to resolve the knee frequency of the differenced noise.
        
        The application of this concept to prototype \Planck-LFI radiometer data has shown
        that the ``in software'' offset balancing method is effective and that the
        balancing condition can be calculated from a limited amount of raw radiometric data in total
        power mode. In particular by methods 1 and 3 applied to a 15~min data stream of
        experimental radiometric data
        it was possible to calculate the {\em same} (at the level of 0.001\%)
        value of $r$.

        Although in \Planck-LFI method 1 is considered as baseline, the other methods
        will be used during the ground data analysis for cross-checks, as they are sensitive
        to non idealities in different ways.
        
        Further studies will be aimed at a better understanding of the impact of
        systematic effects on the gain modulation factor accuracy and to continue the analysis of
        laboratory radiometer data in order to check our predictions. In addition we will continue
        with more realistic and detailed simulations.

\begin{acknowledgements}
        The research described in this paper was performed in part at the Jet Propulsion
        Laboratory, California Institute of Technology, under a contract with the National
        Aeronautics and Space Administration.
\end{acknowledgements}

\bibliographystyle{aa} 
\bibliography{bibliography} 

\end{document}